\documentstyle[twocolumn,eqsecnum,aps]{revtex}

\begin{document}
\draft
\title{Correlated errors in quantum error corrections
 }
\author{W.Y.Hwang \cite{email},
 D.Ahn \cite{byline},
 and S. W. Hwang \cite{byline2}}

\address{ Institute of Quantum Information Processing
and Systems,
 University of Seoul 90, Jeonnong, Tongdaemoon,
 Seoul 130-743, Korea
}
\maketitle
\begin{abstract}
We show that errors are not generated correlatedly provided
that quantum bits do not directly interact with (or couple
to) each other. Generally, this no-qubits-interaction
condition is assumed except for the case where two-qubit
gate operation is being performed. In particular, the
no-qubits-interaction condition is satisfied in the
collective decoherence models. Thus, errors are not
correlated in the collective decoherence. Consequently, we
can say that current quantum error correcting codes which
correct single-qubit-errors will work in most cases
including the collective decoherence.
\end{abstract}

\narrowtext
Information processing with quantum bits (qubits) e.g.
quantum computing and quantum cryptography is a novel
technique that will solve some classically intractable
problems \cite{feyn}-\cite{bene}. However, in order to make
quantum computing practical, quantum error
correcting codes (QECCs) \cite{shor}-\cite{ben2} are
indispensable \cite{pres}. With QECCs,
we can correct errors
on qubits induced by interactions of qubits with
the environment.

However, there exists no QECC that can correct all errors.
That is, only some subsets of all possible errors can be
corrected with QECCs. So, the strategy is to choose certain
subclasses of errors that constitute dominant parts
as to-be-corrected ones, while other classes of errors that
constitute negligible parts as not-to-be-corrected
ones. Generally, single-qubit-errors where only one qubit
has undergone interaction with the environment or arbitrary
unitary operation are assumed to be the most common ones.
More precisely, it
is assumed that {\it the probability of $k$ (integer $k$
$\geq0$) errors is of order $\epsilon^k$},
which is much smaller than $\epsilon$ the
probability of a single error if $\epsilon$ is small enough
and $k\geq2$ \cite{pres}.
This is the independence condition. However, it
should be noted that the independence condition is
distinguished from the independent decoherence where each
qubits interact with their own environments which
do not interact with one another
\footnote{Correlated decoherence should also be
distinguished from collective decoherence.
The former is the one which does
not satisfy the independence condition while the latter is
the one where qubits interacts with environments
collectively.}. Although the independence of
qubit-environment interaction ensures the independence
condition, the converse is not guaranteed.
The purpose of this
paper is to show that even if qubits do not interact
independently with environments, the generated errors
satisfy the independence condition to the second order,
provided that quantum bits do not directly interact
with (or couple to) each other. Generally, this
no-qubits-interaction condition is assumed except for the
case where two-qubit gate operation is being performed. In
particular, the no-qubits-interaction condition is satisfied
in the collective decoherence models
\cite{dua2}-\cite{lida}. Thus, we
can say that correlated errors are not generated in most
cases including the collective decoherence. Therefore,
current QECCs \cite{shor}-\cite{lafl} which correct
single-qubit-errors work in most cases including the
collective decoherence. Recently Knill et
al. have shown that there exist some QECCs that
can correct errors due to general interaction \cite{kni3}.
So, there exist some QECCs which correct
errors due to collective interaction. However, their
results do not mean that QECCs correcting
single-qubit-errors work in collective decoherence.

First, let us consider complete independent
decoherence where qubits interact with their own
environments which do not interact with one another.
This has been addressed and worked out thoroughly
in Ref.\cite{knil,kni2}. We will consider this in
Hamiltonian formulations.
Let us consider the following total Hamiltonian.
\begin{eqnarray}
\label{a}
{\bf H}_T&=&
[{\bf H}_1 \otimes {\bf I}_2 \otimes \cdot\cdot\cdot
\otimes {\bf I}_n \otimes
{\bf I}_1^E \otimes {\bf I}_2^E \otimes \cdot\cdot\cdot
\otimes {\bf I}_n^E
\nonumber\\
&+&
{\bf I}_1 \otimes {\bf I}_2 \otimes \cdot\cdot\cdot
\otimes {\bf I}_n \otimes
{\bf H}_1^E \otimes {\bf I}_2^E \otimes \cdot\cdot\cdot
\otimes {\bf I}_n^E
\nonumber\\
&+&
\sum_{j}{\bf Q}_1^{j} \otimes {\bf I}_2 \otimes
\cdot\cdot\cdot \otimes {\bf I}_n \otimes
{\bf E}_1^j \otimes {\bf I}_2^E \otimes \cdot\cdot\cdot
\otimes {\bf I}_n^E]
\nonumber\\
&+&
[{\bf I}_1 \otimes {\bf H}_2 \otimes \cdot\cdot\cdot
\otimes {\bf I}_n \otimes
{\bf I}_1^E \otimes {\bf I}_2^E \otimes \cdot\cdot\cdot
\otimes {\bf I}_n^E
\nonumber\\
&+&
{\bf I}_1 \otimes {\bf I}_2 \otimes \cdot\cdot\cdot
\otimes {\bf I}_n \otimes
{\bf I}_1^E \otimes {\bf H}_2^E \otimes \cdot\cdot\cdot
\otimes {\bf I}_n^E
\nonumber\\
&+&\sum_{j}{\bf I}_1 \otimes {\bf Q}_2^{j} \otimes
\cdot\cdot\cdot \otimes {\bf I}_n \otimes
{\bf I}_1^E \otimes {\bf E}_2^j \otimes \cdot\cdot\cdot
\otimes {\bf I}_n^E]+ \cdot\cdot\cdot
\nonumber\\
&+&[{\bf I}_1 \otimes {\bf I}_2 \otimes \cdot\cdot\cdot
\otimes {\bf H}_n \otimes
{\bf I}_1^E \otimes {\bf I}_2^E \otimes \cdot\cdot\cdot
\otimes {\bf I}_n^E
\nonumber\\
&+&
{\bf I}_1 \otimes {\bf I}_2 \otimes \cdot\cdot\cdot
\otimes {\bf I}_n \otimes
{\bf I}_1^E \otimes {\bf I}_2^E \otimes \cdot\cdot\cdot
\otimes {\bf H}_n^E
\nonumber\\
&+&\sum_{j}{\bf I}_1 \otimes {\bf I}_2 \otimes
\cdot\cdot\cdot \otimes {\bf Q}_n^j \otimes
{\bf I}_1^E \otimes {\bf I}_2^E \otimes \cdot\cdot\cdot
\otimes {\bf E}_n^j].
\end{eqnarray}
Here, ${\bf H}_{\alpha}$ and ${\bf H}_{\alpha}^E$
are the free
Hamiltonian of $\alpha$-th qubit and $\alpha$-th
environment, respectively,
($\alpha=1,2,...,n$ and $n$ is the number
of qubits and integer $j \geq 1$)
and ${\bf I}$ is the identity operator.
${\bf Q}^j_\alpha$ is an operator
that acts on $\alpha$-th qubit and
${\bf E}_\alpha^{j}$ is an operator that
acts on $\alpha$-th environment.
It is clear that a set of terms in a parenthesis
commute with that in other parenthesis in Eq.(\ref{a}).
Since
$\exp(\hspace{1mm} \sum_{i} A_i)=
\prod_{i} \exp(A_i)$ when $[A_i, A_j]=0$ for each $i,j$
($[A,B]=AB-BA$), the total unitary time evolution operator
$U(t)=\exp(-i{\bf H}_T t)$ decomposes into $n$
factors.
Thus each qubit-environment system
evolves separately by their
own unitary operators, for example, the first qubit-
environment system
by $U_1(t)=\exp(-i[{\bf H}_1 \otimes {\bf I}_2 \otimes \cdot\cdot\cdot
\otimes {\bf I}_n \otimes
{\bf I}_1^E \otimes {\bf I}_2^E \otimes \cdot\cdot\cdot
\otimes {\bf I}_n^E+
{\bf I}_1 \otimes {\bf I}_2 \otimes \cdot\cdot\cdot
\otimes {\bf I}_n \otimes
{\bf H}_1^E \otimes {\bf I}_2^E \otimes \cdot\cdot\cdot
\otimes {\bf I}_n^E
+\sum_{j}{\bf Q}_1^{j} \otimes {\bf I}_2 \otimes
\cdot\cdot\cdot \otimes {\bf I}_n \otimes
{\bf E}_1^j \otimes {\bf I}_2^E \otimes \cdot\cdot\cdot
\otimes {\bf I}_n^E]t)$.
Each qubit-environment's evolution can be decomposed
\cite{shor,pres} as,
for example,
\begin{eqnarray}
\label{b}
U_1(t)
|\psi\rangle_1 |e\rangle_1
&=& \sum_{k=0}^3(\sigma^k \otimes {\bf I}_2 \otimes
\cdot\cdot\cdot \otimes {\bf I}_n)|\psi\rangle_1
|e_k\rangle_1
\nonumber\\
&\equiv& \sum_{k=0}^3 \sigma_1^k
|\psi\rangle_1 |e_k\rangle_1.
\end{eqnarray}
Here, $|\psi\rangle_{\alpha}$ and $|e\rangle_{\alpha}$
denotes  $\alpha$-th qubits and
$\alpha$-th environment state, respectively.
$\sigma^0= {\bf I},
\sigma^1=\sigma^x, \sigma^2= -i\sigma^y$,
and $\sigma^3=\sigma^z$, ${\bf I}$ is the identity operator,
and $\sigma^x, \sigma^y, \sigma^z $
are the Pauli operators.
$\sigma_{\alpha}^k$ denotes $\sigma^k$ acting
on $\alpha$-th
qubit leaving others intact. $|e_k\rangle$ are not
normalized and not necessarily orthogonal \cite{shor,lafl}.
However, in general the norm of the
terms with $\sigma_1^1, \sigma_1^2, \sigma_1^3$ in
Eq.(\ref{b})
are of the first order of time $t$ while that
with $\sigma_1^0$ is of the zeroth order.
This property is required to ensure the quantum
Zeno effect \cite{misr}-\cite{duan}. Therefore,
\begin{equation}
\label{c}
\sum_{k=0}^3 \sigma_1^k |\psi\rangle_1 |e\rangle_1
= c_1^0 t^0 \sigma^0_1|\psi\rangle_1 |\bar{e}_0\rangle_1+
  c_1^k t \sum_{k=1}^3 \sigma_1^k |\psi\rangle_1
  |\bar{e}_k\rangle_1,
\end{equation}
where $|\bar{e}_k\rangle$ is normalized state of
$|e_k\rangle$ and $c_1^k$'s are some constants.
The same relation is satisfied for other $\alpha$'s.
As noted above, the total qubits-environments system can be
expressed as direct products of each qubit-environment
system, each of which satisfy an equation similar to
Eq.(\ref{c}). Then, we can see by inspection that
{\it terms with $k$ errors are of order $t^k$}
in general (Note that the
total state is in a form similar
to $[1+t]^n$.). So we can say that the independence of
qubits-environments interactions ensure the independence
condition.

Next,
let us consider incomplete independent decoherence where
qubits interact with different environments which
are still interacting with one another.
In this case total states do not
decompose into factors in general and thus the above
method
cannot be used to derive the independence condition.
On the other hand,
one may guess that collective decoherence generates
correlated errors. However, there is no reason why the
collective interaction of qubits with the environment
necessarily induces correlated errors.
However, in both models, qubits do not couple to each
other or they satisfy no-qubits-interaction condition.
Then correlated errors are not generated,
as we show in the following.
Therefore, we can say that incomplete and
collective decoherence does not generate correlated errors.
Now, we state the no-qubits-interaction condition more
precisely: in each term of the qubit-environment interaction
Hamiltonian ${\bf H}_I$, only one qubit-operator is a
non-identity. That is,
\begin{eqnarray}
\label{d}
{\bf H}_I&=&
\sum_{j}{\bf Q}_1^{j} \otimes {\bf I}_2 \otimes
\cdot\cdot\cdot \otimes {\bf I}_n \otimes {\bf E}_1^{j}
\nonumber\\
&+&
\sum_{j} {\bf I}_1 \otimes  {\bf Q}_2^{j} \otimes
\cdot\cdot\cdot \otimes {\bf I}_n \otimes {\bf E}_2^{j}
\nonumber\\
&+&
\cdot\cdot\cdot+
\sum_{j} {\bf I}_1 \otimes {\bf I}_2 \otimes \cdot\cdot\cdot
\otimes  {\bf Q}_n^{j} \otimes {\bf E}_n^{j}.
\end{eqnarray}
The total Hamiltonian is the following.
\begin{eqnarray}
\label{e}
{\bf H}_T&=&
{\bf H}_1 \otimes {\bf I}_2 \otimes \cdot\cdot\cdot
\otimes {\bf I}_n \otimes {\bf I}_E
+
{\bf I}_1 \otimes {\bf H}_2 \otimes \cdot\cdot\cdot
\otimes {\bf I}_n \otimes {\bf I}_E
\nonumber\\
&+& \cdot\cdot\cdot +
{\bf I}_1 \otimes {\bf I}_2 \otimes \cdot\cdot\cdot
\otimes {\bf H}_n \otimes {\bf I}_E
\nonumber\\
&+&
{\bf I}_1 \otimes {\bf I}_2 \otimes \cdot\cdot\cdot\otimes
{\bf I}_n \otimes {\bf H}_E
+{\bf H}_I
\equiv
{\bf H}_0+ {\bf H}_I.
\end{eqnarray}
Here we adopt the interaction picture
\cite{saku} where $|\psi\rangle_I$ (the state vector in the
interaction picture) $= \exp(it{\bf H}_0)
|\psi\rangle_S$ (the state vector in Schrodinger picture).
The time evolution of $|\psi\rangle_I$ is determined by
the Schrodinger-like equation
\begin{equation}
\label{f}
i\frac{\partial{|\psi\rangle_I}}{\partial t}=
V(t) \hspace{1mm} |\psi\rangle_I,
\end{equation}
where
\begin{equation}
\label{g}
V(t)\equiv \exp(it{\bf H}_0) \hspace{1mm} {\bf H}_I
\exp(-it{\bf H}_0).
\end{equation}
Since $V(t)$ is time dependent, the time evolution operator
$U_I(t)$ for $|\psi\rangle_I$ is given by the Dyson series
\cite{saku}.
\begin{eqnarray}
\label{h}
&&U_I(t) =1+
\nonumber\\
&&\sum_{m=1}^{\infty} (-i)^m
\int^t_0 dt_1 \int^{t_1}_0 dt_2 \cdot\cdot\cdot
\int^{t_{m-1}}_0 dt_m V(t_1)V(t_2)\cdot\cdot\cdot V(t_m).
\nonumber\\
\end{eqnarray}
From Eqs.(\ref{d}) and (\ref{g}),
\begin{eqnarray}
\label{i}
V(t)&=&
\exp(it{\bf H}_0)[\sum_{j}{\bf Q}_1^{j} \otimes {\bf I}_2
\otimes \cdot\cdot\cdot \otimes {\bf I}_n \otimes {\bf
E}_1^{j}] \exp(-it{\bf H}_0)
\nonumber\\
&+&
\exp(it{\bf H}_0)[\sum_{j} {\bf I}_1 \otimes  {\bf Q}_2^{j}
\otimes \cdot\cdot\cdot \otimes {\bf I}_n \otimes {\bf
E}_2^{j}]\exp(-it{\bf H}_0)
\nonumber\\
+\cdot\cdot\cdot &+&
\exp(it{\bf H}_0)[\sum_{j} {\bf I}_1 \otimes {\bf I}_2
\otimes \cdot\cdot\cdot \otimes {\bf Q}_n^{j}
\otimes {\bf E}_n^{j}] \exp(-it{\bf H}_0)
\nonumber\\
&\equiv&
V_1(t)+ V_2(t)+ \cdot\cdot\cdot+ V_n(t).
\end{eqnarray}
We consider the relation
\begin{equation}
\label{j}
U_I(t)= U_I^1(t) U_I^2(t) \cdot\cdot\cdot U_I^n(t) +
       O(t^2),
\end{equation}
where $U_I^{\alpha}(t)= 1+ \sum_{m=1}^{\infty} (-i)^m
\int^t_0 dt_1 \int^{t_1}_0 dt_2 \cdot\cdot\cdot
\int^{t_{m-1}}_0 dt_m V_{\alpha}(t_1) V_{\alpha}(t_2)
\cdot\cdot\cdot V_{\alpha}(t_m)$ and $O(f(x))$ means
asymptotically less than a constant operator times $f(x)$.
However, since $|\psi\rangle_S= \exp(-it{\bf
H}_0)|\psi\rangle_I$ and the operator $\exp(-it{\bf H}_0)$
do not entangle qubits with environments, it is
sufficient for us to consider only $U_I(t)$.
We can see that each $U_I^{\alpha}(t)$ makes
$\alpha$-th qubit to entangle with environment.
For an example,
\begin{eqnarray}
\label{k}
U_I^1(t)
|\psi\rangle_I |e\rangle_I
&=& \sum_{k=0}^3(\sigma^k \otimes {\bf I}_2 \otimes
\cdot\cdot\cdot \otimes {\bf I}_n)|\psi\rangle_I
|e_k\rangle_I
\nonumber\\
&\equiv&
\sum_{k=0}^3 \sigma_1^k |\psi\rangle_I |e_k\rangle_I.
\end{eqnarray}
Here, $|\psi\rangle_I$ and $|e\rangle_I$ denotes  qubits and
the environment state in the interaction picture,
respectively. And $|e_k\rangle_I$ are not
normalized and not necessarily orthogonal.
By operating all factors in $U_I(t)$ sequentially,
we obtain
\begin{eqnarray}
\label{l}
U_I(t)|\psi\rangle_I |e\rangle_I
&=&
\sum_{\{k\}}
\sigma_1^{k_1}\sigma_2^{k_2}\cdot\cdot\cdot \sigma_n^{k_n}
|\psi\rangle_I |e_{\{k\}}\rangle_I
\nonumber\\
&+& O(t^2)|\psi\rangle_I |e\rangle_I,
\end{eqnarray}
where $\{k\}$ is an abbreviation for $k_1,k_2,...,k_n$, and
$k_{\alpha}= 0,1,2,3$.
Let us consider Eq.(\ref{k}). As above, the norm of the
terms with $\sigma_1^1, \sigma_1^2, \sigma_1^3$ of
Eq.(\ref{k})
are of the first order of time $t$ while the norm of the
term with $\sigma_1^0$ is of the zeroth order of time $t$.
Therefore,
\begin{eqnarray}
\label{m}
\sum_{k=0}^3 \sigma_1^k |\psi\rangle_I |e\rangle_I
&=&
c_1^0 t^0 \sigma^0_1|\psi\rangle_I |\bar{e}_0\rangle_I+
c_1^k t \sum_{k=1}^3 \sigma_1^k |\psi\rangle_I
|\bar{e}_k\rangle_I
\nonumber\\
&+& O(t^2)|\psi\rangle_I |e\rangle_I,
\end{eqnarray}
where $|\bar{e}_k\rangle_I$ is the normalized state of
$|e_k\rangle_I$.
The same relation is satisfied for other $\alpha$'s.
Then,
\begin{eqnarray}
\label{n}
&&
\sum_{\{k\}}
\sigma_1^{k_1}\sigma_2^{k_2}\cdot\cdot\cdot \sigma_n^{k_n}
|\psi \rangle_I |e_{\{k\}}\rangle_I
+ O(t^2)|\psi\rangle_I |e\rangle_I
\nonumber\\
&=&
\sum_{\{k\}}
c_{\{k\}} t^{N(\{k\})}
\sigma_1^{k_1}\sigma_2^{k_2}\cdot\cdot\cdot \sigma_n^{k_n}
|\psi \rangle_I |\bar{e}_{\{k\}}\rangle_I
\nonumber\\
&+& O(t^2)|\psi\rangle_I |e\rangle_I,
\end{eqnarray}
where $N(\{k\})$ is the number of instances
when $k_{\alpha} \neq 0$.
Now, we can see that {\it all terms with more than
1 errors
(or $N(\{k\}) \geq 2$) are of order $t^2$}. Thus the
independence condition is satisfied to the second order (we
can obtain the full independence condition in the case where
the $O(t^2)|\psi\rangle_I |e\rangle_I$ term is negligible.).
So, we can say that any qubit-environment system that
satisfies the no-qubits-interaction condition
(Eq.(\ref{d})) obeys
the independence condition to the second order so that the
QECCs correcting single-qubit-errors works successfully.

To summarize, we have shown that errors are not
generated correlatedly, provided that quantum bits do not
directly interact with each other, or that in each term of
the qubit-environment interaction Hamiltonian ${\bf H}_I$
only one qubit-operator is a non-identity operator
(Eq.(\ref{d})).
Generally, this no-qubits-interaction condition is assumed
except for the case where two-qubit gate operation is being
performed. In particular, the no-qubits-interaction
condition is satisfied in the collective decoherence models
\cite{dua2}-\cite{lida}.
So, current QECCs \cite{shor}-\cite{lafl}
which correct single-qubit-errors work in most cases
including the collective decoherence.

\acknowledgments
This work was supported by the Korean Ministry of Science
and Technology through the Creative Research Initiatives
Program under Contract No. 98-CR-01-01-A-20.
We are very grateful to Andi Song for corrections of this
paper.

\end{document}